# First-Principles Prediction of Two-Dimensional $B_3C_2P_3$ and $B_2C_4P_2$: Structural Stability, Fundamental Properties, and Renewable Energy Applications

Andrey A. Kistanov[1,*], Stepan A. Shcherbinin[2,3*], Svetlana V. Ustiuzhanina[4], Marko Huttula[1], Wei Cao[1], Vladimir R. Nikitenko[5], and Oleg V. Prezhdo[6]

[1]Nano and Molecular Systems Research Unit, University of Oulu, Oulu 90014, Finland

[2]Peter the Great Saint Petersburg Polytechnical University, Saint Petersburg 195251, Russia

[3]Southern Federal University, Rostov-on-Don 344006, Russia

[4]Institute for Metals Superplasticity Problems Russian Academy of Sciences, Ufa 450001, Russia

[5]National Research Nuclear University MEPhI, Moscow 115409, Russia

[6]Department of Chemistry, University of Southern California, Los Angeles, CA 90089, United States

[*]*Email*: andrey.kistanov@oulu.fi (AAK), stefanshcherbinin@gmail.com (SASh)

**ABSTRACT:** The existence of two novel hybrid two-dimensional (2D) monolayers, 2D $B_3C_2P_3$ and 2D $B_2C_4P_2$, has been predicted based on the density functional theory calculations. It has been shown that these materials possess structural and thermodynamic stability. 2D $B_3C_2P_3$ is a moderate band gap semiconductor, while 2D $B_2C_4P_2$ is a zero band gap semiconductor. It has also been shown that 2D $B_3C_2P_3$ has a highly tunable band gap under the effect of strain and substrate engineering. Moreover, 2D $B_3C_2P_3$ produces low barriers for dissociation of water and hydrogen molecules on its surface, and shows fast recovery after desorption of the molecules. The novel materials can be fabricated by carbon doping of boron phosphide, and directly by arc discharge and laser ablation and vaporization. Applications of 2D $B_3C_2P_3$ in renewable energy and straintronic nanodevices have been proposed.

**TOC:** Novel 2D $B_3C_2P_3$ is predicted and proposed for application in renewable energy devices.

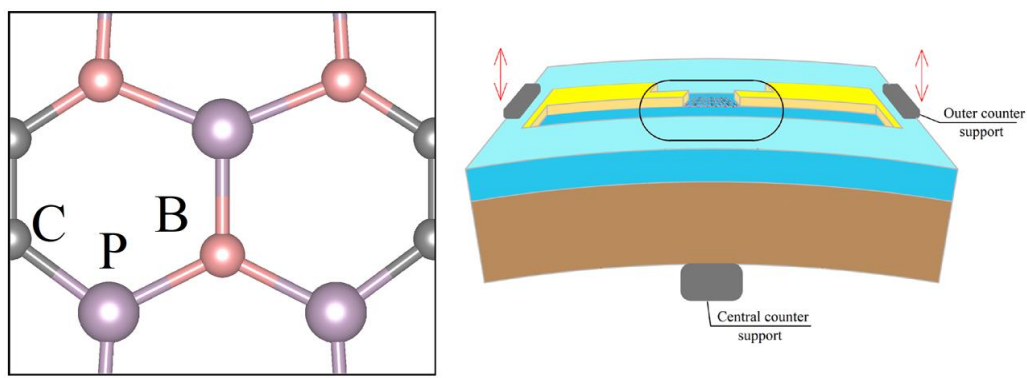

# INTRODUCTION

The discovery of the first two-dimensional (2D) material, graphene, a single atom thick layer of carbon, which is stronger than diamond but stretches like rubber,[1] shook the scientific world. This invention has rendered research on developing, synthesis, and application of 2D materials[2] and has led to breakthroughs in materials science, physics, chemistry, etc. The next reached height in the investigation of 2D materials was a successful development and fabrication of hybrid 2D materials consisting of several elements.[3-5] Most of such multi-element hybrids have shown functional properties superior to these of their individual compounds. For instance, a monolayer boron carbide possesses environmental stability and high chemical activity towards toxic gases.[6] Stable semiconducting boron–graphdiyne and $GeP_3$ monolayers obtained by hybrid constructions of carbon with boron[7] and phosphorus with germanium[8] have controllable band gaps and high carrier mobilities. Another notable example is 2D phosphorus carbide which combines the advantages of its compounds, such as ambient stability,[9] wide and tunable band gap, [10] robust conductivity,[11] and extra-high carrier mobility.[12]

Accordingly, new databases[13,14] containing thousands of novel 2D materials are created and updated daily. The establishment of such databases has become a new milestone in the development of materials, including 2D materials. These databases constitute the building blocks for developing machine learning algorithms, which are becoming a powerful tool in predicting exotic materials with required properties.[15] For instance, hexagonal NaCl thin films have been fabricated based on the ab initio prediction via an evolutionary algorithm.[16] Machine learning algorithms have been capable to predict electronic structure of hybrids of graphene and h-boron nitride with arbitrary supercell configurations.[17] The combination of machine learning and ab initio calculations has predicted key properties of point defects in 2D materials.[18] Furthermore, the presence of extensive databases of 2D materials also simplifies the fabrication and sorting according to certain properties of new 2D materials.[13] Therefore, not only the development of the machine learning algorithms and experimental approaches but also the updates of materials databases with new elements is a challenge for the present.

Currently, hybrid 2D materials are mainly created of elements such as boron, phosphorus, and carbon. Different combinations of these elements allow the production of functional, cheap, and relatively easily fabricated 2D materials. This work presents theoretical prediction of new 2D materials, 2D $B_3C_2P_3$ and 2D $B_2C_4P_2$, based on the density functional theory (DFT) calculations. The structural and thermodynamic stability together with the electronic and mechanical properties of these materials are investigated. In addition, the effect of strain and substrate engineering on the electronic structure and catalytic activity of 2D $B_3C_2P_3$ is considered. Based on the obtained results, several topical applications of 2D $B_3C_2P_3$ are proposed.

## RESULTS

The structure of monolayer $B_xC_yP_z$ is initially designed based on the geometry of the 2D phosphorus carbide.[9] The design of 2D $B_xC_yP_z$ consisted of a systematical substitution of carbon and/or phosphorus atoms with boron atoms. For the unit cell of each obtained structure, a lattice optimization was performed, and the stability of those structures was checked by calculating phonon dispersion spectra and/or by conducting ab initio molecular dynamics (AIMD) calculations. Based on those simulations two stable modifications 2D $B_3C_2P_3$ and 2D $B_2C_4P_2$ were selected. The main attention is focused on 2D $B_3C_2P_3$, while the discussion on the structure and properties of 2D $B_2C_4P_2$ (Figure S1) can be found Supporting Information (SI).

The top and side views of the unit cell of 2D $B_3C_2P_3$ are presented in Figure 1a. According to Figure 1a, a unit cell of 2D $B_3C_2P_3$ includes eight atoms, i.e., three boron atoms, three phosphorus atoms, and two carbon atoms. 2D $B_3C_2P_3$ stabilizes in a 2D honeycomb lattice with the space group 25 *Pmm*2 and the lattice parameters of $a = 6.06$ Å and $b = 5.19$ Å. A close look at 2D $B_3C_2P_3$ reveals that it is formed by strips consisting of boron–phosphorus atoms which are connected by chains of carbon atoms. Such a unique structural order of 2D $B_3C_2P_3$ makes its fabrication possible based on carbon doping of boron phosphide[19] following the carbon doping technique for the synthesis of 2D phosphorus carbide.[3] In addition, BCN and other types of nanotubes have been synthesized using electrical arc discharge, laser ablation, and laser vaporization techniques, which also motivates the synthesis of 2D $B_3C_2P_3$.[20] More information on the design of the structure and structural parameters of 2D $B_3C_2P_3$ and 2D $B_2C_4P_2$, including bond lengths and bond angles, are collected in Section 2 in SI.

The electronic localization function (ELF) reflects the degree of charge localization in the real space, where 0 represents a free electronic state while 1 represents a perfect localization. The calculated ELF for 2D $B_3C_2P_3$ (Figure 1b) reflects electron density shared by neighboring atoms, which emphasizes the covalent character of bonding in 2D $B_3C_2P_3$ and suggests the stability of the material. Since the electronegativity of C is greater than that of P, while the electronegativity of P is greater than that of B, the P-B bond is the strongest covalent bond. The kinetic stability of 2D $B_3C_2P_3$ is also confirmed by calculating the phonon dispersion spectra along the high symmetry directions ($\Gamma \rightarrow X \rightarrow S \rightarrow Y \rightarrow \Gamma$) of the Brillouin zone (Figure 1c). It should be noted that the transverse acoustic (TA), longitudinal acoustic (LA), and the out-of-plane z-direction acoustic (ZA) modes display the normal linear dispersion around the $\Gamma$ point. To check the thermal stability of 2D $B_3C_2P_3$, AIMD simulations are performed at 300 K for the 2D $B_3C_2P_3$ structure consisting of 3×3×1 unit cells. The performed calculations show that the 2D $B_3C_2P_3$ structure remains stable after 5 ps (Figure S2 and movie 1 in SI).

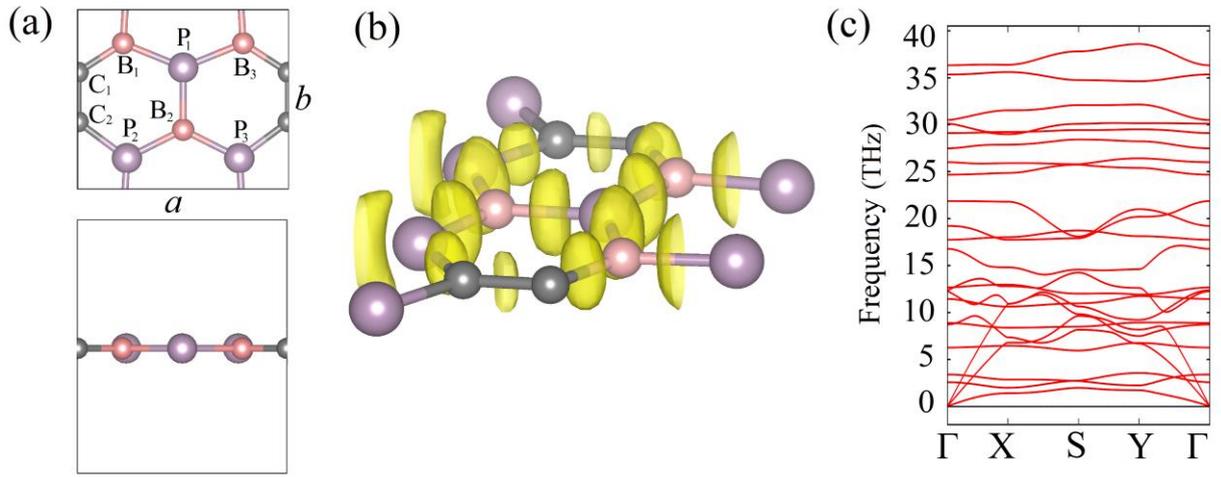

**Figure 1.** (a) Atomic structure, (b) ELF, and (c) phonon dispersion curves for the 2D $B_3C_2P_3$ unit cell. The boron, phosphorus, and carbon atoms are colored in pink, violet, and grey.

The band structure of 2D $B_3C_2P_3$ obtained using both the Perdew-Burke-Ernzerhof (PBE) exchange-correlation functional under the generalized gradient approximation (GGA) and the Heyd–Scuseria–Ernzerhof (HSE06) functional is plotted in Figure 2a. The PBE GGA approach predicts 2D $B_3C_2P_3$ to be an indirect band gap semiconductor with the band gap of 0.35 eV. The HSE06 approach also predicts 2D $B_3C_2P_3$ to be an indirect band gap semiconductor, but with a larger band gap of 0.58 eV. According to both calculations, the conduction band minimum (CBM) is located at the vicinity of the X point, while the valence band maximum (VBM) is located between the X and Γ points. The partial density of states (PDOS) of 2D $B_3C_2P_3$ (Figure 2b) demonstrates that $p_z$ states of phosphorus atoms give the main contribution to the VBM, while the CBM forms due to a strong mixing of $p_z$ states of boron, carbon, and phosphorus atoms. The major impact of phosphorus atoms on VBM, as well as orbital mixing in CBM, can be attributed to the difference of the length of the P-B and P-C bonds in 2D $B_3C_2P_3$. It is the result of strong interaction of phosphorus atoms ($P_2$ and $P_3$ in Figure 1a), as indicated by the charge localization at the P-P bonds in ELF (Figure 1b). Such a distortion of phosphorus bonds leads to a deformed honeycomb lattice of 2D $B_3C_2P_3$ and determines its electronic structure. A similar impact of lattice distortions on the electronic structure has been observed in phosphorene.[21]

Figure 2c shows the energy diagram of the workfunction of 2D $B_3C_2P_3$ in comparison with other 2D materials and bulk metals possessing high workfunctions. The workfunction of 2D $B_3C_2P_3$ is 4.72 eV, which is slightly higher than that of graphene and phosphorene,[22,23] but lower than that of 2D PC[24] or borophene.[25] The relatively low workfunction of 2D $B_3C_2P_3$ can be attributed to the nature of its atomic states around the Fermi level consisting of the out-of-plane $p_z$ states (this is similar e.g. to graphene), which lie above the in-plane $s$-$p$ hybridized states as, for example, in borophene. Thus, the ionization of 2D $B_3C_2P_3$ is comparable to that in graphene, while it is lower than that in borophene.

The frequency-dependent dielectric function of 2D $B_3C_2P_3$ is calculated to show its optical absorption characteristics. The material presents different adsorption characteristics in the *x*, *y*, and *z* directions (Figure S3 in SI), which can be attributed to its high anisotropy, similarly to the case of phosphorene[26] (more information is presented in section 4 in SI). The calculated Young's moduli of 2D $B_3C_2P_3$ along the armchair and zigzag directions are equal to 376 GPa and 365 GPa, respectively. The obtained Young's moduli are smaller than those of graphene[27] but comparable to that of α-PC along the zigzag direction (348.69 GPa).[28] The shear modulus of 2D $B_3C_2P_3$ reaches 134 GPa, and the Poisson's ratios are 0.24 and 0.25 for tension along the armchair and zigzag directions, respectively.

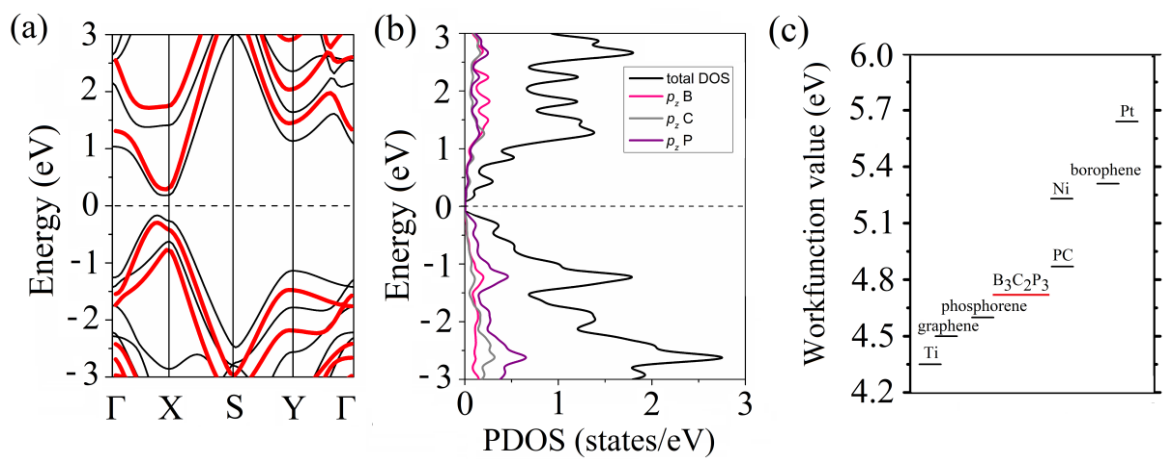

**Figure 2.** (a) Band structure of 2D $B_3C_2P_3$. The black and red lines show the band structure calculated by the PBE and HSE approaches. (b) PDOS of 2D $B_3C_2P_3$ (calculated by the PBE). (c) Comparison of the workfunction of $B_3C_2P_3$ with those of common 2D materials and bulk metals.

2D $B_3C_2P_3$ consists of several chemically active elements. Hence, it is intriguing to consider its applications in gas sensing, photovoltaic, and energy storage devices. We investigate interaction of 2D $B_3C_2P_3$ with water ($H_2O$) and hydrogen ($H_2$) molecules. Firstly, the adsorption behavior of $H_2O$ and $H_2$ on 2D $B_3C_2P_3$ is studied. For that, all possible configurations and adsorbing sites are considered with the $H_2O$ and $H_2$ molecules aligned parallel, perpendicular or tilted to the 2D $B_3C_2P_3$ surface. The determined lowest-energy configuration for the $H_2O$ molecule on 2D $B_3C_2P_3$ is shown in Figure S4a. The $H_2O$ molecule is located at the center of the hexagon with two H atoms directed to the 2D $B_3C_2P_3$ surface. The distance between the $H_2O$ molecule and the 2D $B_3C_2P_3$ surface is found to be 2.59 Å. The adsorption energy $E_a$ of $H_2O$ on 2D $B_3C_2P_3$ is -0.15 eV which is comparable to these in the most chemically active 2D materials such as InSe,[29] graphene,[30,31] and pnictogens.[32] The differential charge density (DCD) plot (Figure S4a) together with the charge transfer analysis (Figure S4a) show an accumulation of electrons in the $H_2O$ molecule (acceptor to 2D $B_3C_2P_3$) with a total charge transfer of 0.023 *e* per molecule. The

PDOS plot (Figure S4b) for the $H_2O$ molecule adsorbed on 2D $B_3C_2P_3$ indicates the absence of $H_2O$-originated states in the vicinity of the VBM and CBM of the host 2D $B_3C_2P_3$. However, the $1b_2$, $3a_1$, and $1b_1$ orbitals of the $H_2O$ molecule are broadened and coincide with the valence states of 2D $B_3C_2P_3$. Therefore, the above-listed results suggest strong affinity of 2D $B_3C_2P_3$ to $H_2O$.

The $H_2$ molecule is 2.63 Å above the 2D $B_3C_2P_3$ surface in its lowest-energy configuration, with the two H atoms located above two face-by-face carbon atoms in the same hexagon, as shown in Figure S4c. The adsorption energy of $H_2$ on 2D $B_3C_2P_3$ is -0.05 eV which is higher than for $H_2$ adsorbed on graphene[30] and phosphorene,[33] but similar to $H_2$ adsorbed on InSe[29] and pnictogens.[32] The DCD plot (Figure S4c) and the charge transfer analysis (Figure S4c) revealed a thin electron transfer from the 2D $B_3C_2P_3$ surface to the $H_2$ molecule (acceptor to 2D $B_3C_2P_3$) with a total charge transfer of 0.008 $e$ per molecule. The sharp peak in the PDOS plot (Figure S4d) of the $H_2$ molecule adsorbed on 2D $B_3C_2P_3$ indicates its relatively weak interaction with the 2D $B_3C_2P_3$ surface.

In both cases, the molecular bond lengths elongate upon adsorption of the $H_2O$ and $H_2$ molecules on 2D $B_3C_2P_3$. The O-H bonds of $H_2O$ elongate from 0.96 Å to 0.98 Å, while the H-H bond of $H_2$ elongates from 0.74 Å to 0.75 Å upon adsorption on 2D $B_3C_2P_3$. Such elongation should facilitate dissociation of $H_2O$ and $H_2$ on 2D $B_3C_2P_3$. Therefore, the energy barrier $E_b$ for the dissociation of $H_2O$ and $H_2$ on 2D $B_3C_2P_3$ is further calculated. The detailed pathways from the initial state (IS) to the transition state (TS) and to the final state (FS) together with the dissociative adsorption reaction barriers for $H_2O$ and $H_2$ on 2D $B_3C_2P_3$ are shown in Figure 3a-d. For the $H_2O$ case, the calculated $E_b$ is 0.98 eV which is lower than that on phosphorene[21,34] and comparable with that on graphene[35] and modified $MoS_2$.[36] For the $H_2$ case, the calculated $E_b$ is 1.01 eV, which is several times lower than that on pure and modified graphene[37] and comparable to that on phosphorene.[38,39] The recovery time is a critical characteristic for the application of the material in hydrogen storage devices.[40,41] According to the conventional transition state theory, 2D $B_3C_2P_3$ possesses fast recovery time of ~7 fs at the room temperature of 300 K.

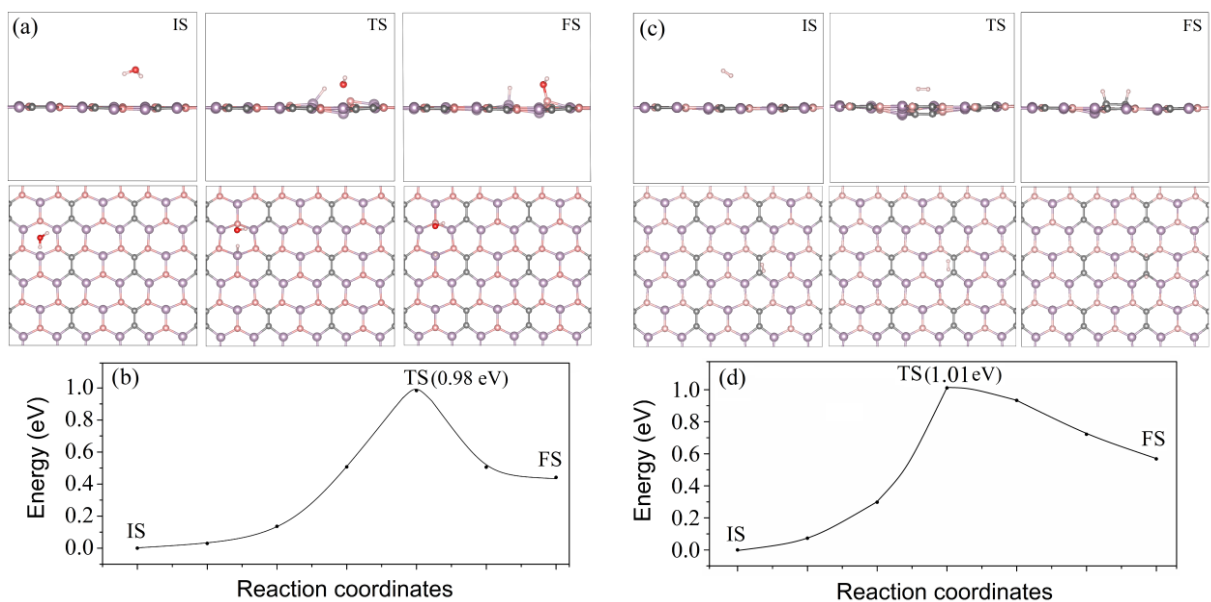

**Figure 3.** The atomic configurations and energy profiles of the reaction pathway from the physisorbed (IS) to the chemisorbed (FS) state in the dissociation process of (a) and (b) $H_2O$ and (c) and (d) $H_2$ molecules on 2D $B_3C_2P_3$.

Previous studies showed that $O_2$ may oxidize chemically active 2D materials such as phosphorene.[42] Our calculations predicted that $O_2$ has a comparably low $E_a = -0.83$ eV on 2D $B_3C_2P_3$. Therefore, 2D $B_3C_2P_3$ may possess high affinity to $O_2$, and future investigations should be targeted to the examination of chemical reaction of $O_2$ with 2D $B_3C_2P_3$.

For the potential application in photovoltaic and energy storage devices the material should also possess specific electronic structure. Basic requirements for photocatalytic materials for hydrogen and water splitting is that the band edges should straddle the hydrogen and water redox potentials.[43] Therefore, the band gap size as well as the tunability of the band gap are essential characteristics of photocatalytic materials.[44,45] The possibility of controlling the band gap size of 2D $B_3C_2P_3$ and ways for the realization of this process are considered. For that, strain engineering which is one of the most functional methods[46] for an engineering of band gap of 2D materials is implemented. Figure 4a shows the band gap size of 2D $B_3C_2P_3$ as a function of applied axial compressive and tensile strains. Gradual increase of an axial compressive strain from 0% to 5% leads to a decrease of the band gap of 2D $B_3C_2P_3$ from 0.35 eV (PBE GGA value) to 0.24 eV (PBE GGA value). In turn, an increase of an axial tensile strain leads to an increase of the band gap of 2D $B_3C_2P_3$ from 0.35 eV (PBE GGA value) to 0.46 eV (PBE GGA value). Note, further increase of the tensile stain leads to a further increase of the band gap of 2D $B_3C_2P_3$ (not shown here). Figure 4b show a schematic of the integration of 2D $B_3C_2P_3$ and straintronics in typical nanoelectronic devices.

Further, the band edge positions for the 2D $B_3C_2P_3$ under applied strains with respect to hydrogen redox potentials are depicted in Figure S5. The VBM of pure 2D $B_3C_2P_3$ is located above the oxidation potential of $O_2/H_2O$, and the CBM is located below the reduction potential of $H^+/H_2$. However, under the small compressive strains of 3% (and larger) the CBM is located above the reduction potential of $H^+/H_2$. Hence, photocatalytic characteristics of 2D $B_3C_2P_3$ may be significantly improved by strain engineering. Furthermore, various substrates such as $MoS_2$ can be actively applied for boosting the catalytic activity of 2D materials.[47,48] In this work, band edge positions of pure 2D $B_3C_2P_3$ and 2D $B_3C_2P_3$ under compressive strain of 5% are analyzed and compared with these of 2D $B_3C_2P_3$ on the $MoS_2$ substrate and with these of pure $MoS_2$. According to Figure 4c, the VBM and CBM of the 2D $B_3C_2P_3/MoS_2$ system are lower than those of pure 2D $B_3C_2P_3$, which suggests type-II band alignment in the 2D $B_3C_2P_3/MoS_2$ heterostructure (for more details see Figure S6 in SI). Previously it has been shown that 2D heterostructures with type-II band alignments can efficiently stimulate the transfer of electron-hole pairs in different layers and reduce the recombination rate.[49] The presence of the $MoS_2$ substrate also leads to a modification of the electronic structure of 2D $B_3C_2P_3$ so that its VBM shifts below the oxidation potential of $O_2/H_2O$. In turn, there is type-I band alignment between pure $MoS_2$ and the 2D $B_3C_2P_3/MoS_2$ system. Though the structure of 2D $B_3C_2P_3$ and its stability have been established computationally, the properties are yet to be explored in other emerging domains concerning its quasi-metallic band types. For example, the cross-over of the VB and CB assembles graphene's band structure, enabling applications similar to those of graphene.

## DISCUSSION

In summary, DFT-based calculations have been utilized to propose two novel hybrid 2D materials, 2D $B_3C_2P_3$ and 2D $B_2C_4P_2$. Among these two, 2D $B_3C_2P_3$ possesses exceptional electronic and mechanical properties. The band structure of 2D $B_3C_2P_3$ with a direct and flexible bandgap resembles that of $MoS_2$, making it promising in many aspects including the application in layered logic storage.[50] The materials can be synthesized by carbon doping of boron phosphide, as well as directly by laser ablation and vaporization, and electrical arc discharge. Furthermore, the observed tunability of the 2D $B_3C_2P_3$ band gap and the demonstrated enhancement of its photovoltaic performance by strain and substrate engineering suggest broad prospects for the application of 2D $B_3C_2P_3$ in straintronic and renewable energy devices.[51,52]

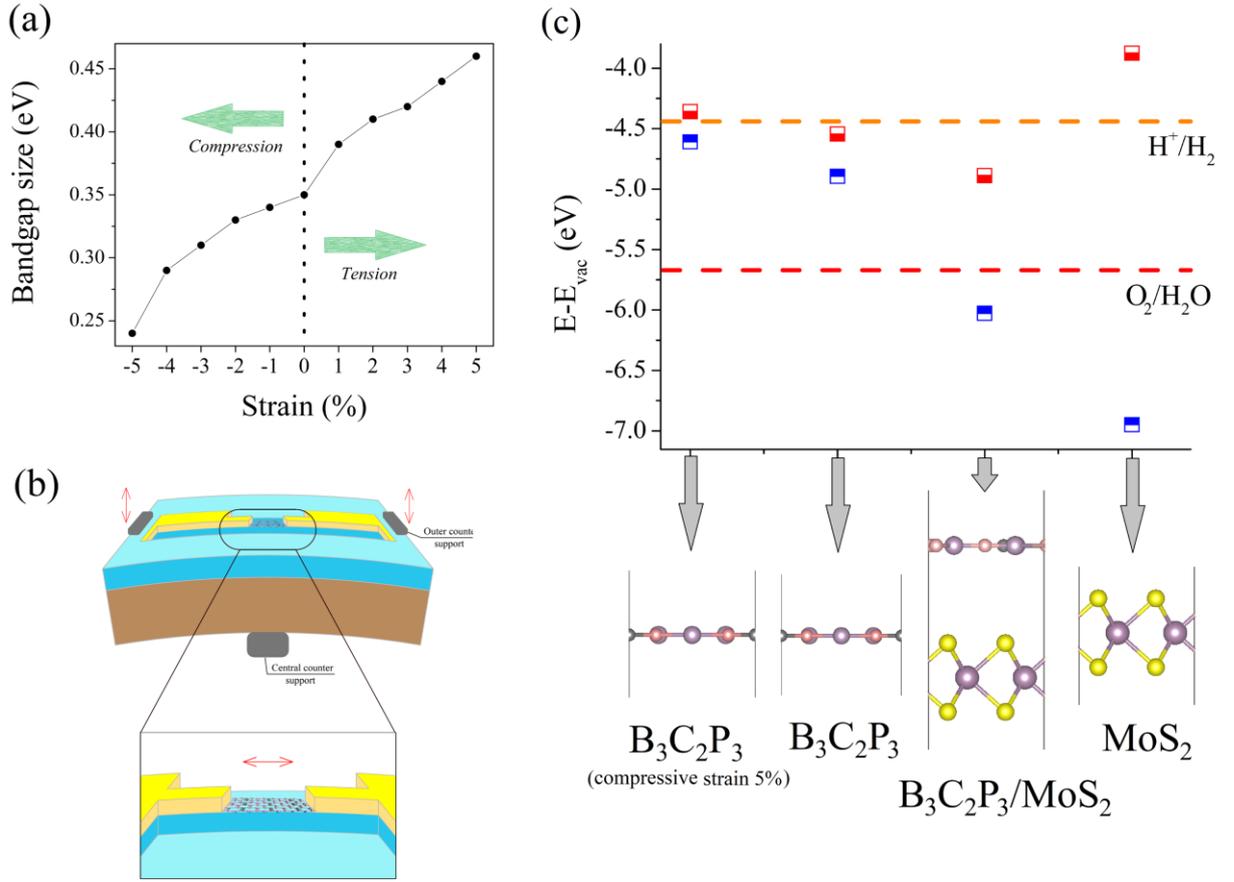

**Figure 4.** (a) The band gap of 2D $B_3C_2P_3$ as a function of applied strain. (b) A schematic of the integration of 2D $B_3C_2P_3$ and straintronics in typical nanoelectronic devices. (c) The band edge positions of pure 2D $B_3C_2P_3$, 2D $B_3C_2P_3$ under compressive strain of 5%, 2D $B_3C_2P_3$ on the $MoS_2$ substrate, and pure $MoS_2$. The hydrogen reduction potential ($H^+/H_2$) and the oxidation potential of $O_2/H_2O$ are represented by the orange and red dashed lines.

## METHODS

The simulations were performed within the framework of spin-polarized DFT as implemented in Vienna Ab Initio Simulation Package (VASP).[53] The PBE exchange-correlation functional under GGA was used for the geometry optimization.[54] All the structures were fully optimized until the atomic forces became smaller than 0.01 eV/Å. The 10×10×5 *k*-mesh and the plane-wave cut-off energy of 450 eV were used. For describing electronic interactions, the HSE06 functional[55] was used in the band structure calculations. The noncovalent chemical interactions between 2D $B_3C_2P_3$ and $H_2O$ and $H_2$ molecules were described using van der Waals-corrected functional with Becke88 optimization (optB88).[56] The adsorption energy $E_a$ of a molecule on 2D $B_2C_4P_2$ was calculated as

$$E_a = E_{tot} - E_m - E_{mol}, \qquad (1)$$

where $E_{tot}$, $E_m$, and $E_{mol}$ are the energies of the molecule-adsorbed 2D $B_3C_2P_3$, the isolated 2D $B_3C_2P_3$, and the molecule.

The charge transfer analysis was conducted by the calculation of the differential charge density (DCD) $\Delta\rho(r)$, which is defined as

$$\Delta\rho(r) = \rho_{tot}(r) - \rho_m(r) - \rho_{mol}(r), \tag{2}$$

where $\rho_{tot}(r)$, $\rho_m(r)$, and $\rho_{mol}(r)$ are the charge densities of the molecule-adsorbed 2D $B_3C_2P_3$, the isolated 2D $B_3C_2P_3$, and the molecule. The exact amount of the charge transfer between the molecule and the surface was calculated by the Bader analysis[57] and by integrating $\Delta\rho(r)$ over the basal plane at the $z$ point for deriving the plane-averaged DCD $\Delta\rho(z)$ along the normal direction $z$ of the sheet.

The reaction barrier was estimated using the climbing image nudged elastic band method.[58] AIMD simulations were performed at the room temperature of 300 K using the Nose-Hoover method with a time step of 1.0 fs.

The compressive strain was defined as

$$\varepsilon = \frac{l-l_0}{l_0} \tag{3}$$

where $l$ and $l_0$ are the lattice constants of the strained and initial supercells, respectively. The Young's modulus of 2D $B_3C_2P_3$ was calculated with the aid of the stress-strain relation.[59]

The recovery time $t$ of 2D $B_3C_2P_3$ was calculated as

$$t = \nu^{-1} e^{\frac{-E_a}{k_B T}} \tag{4}$$

where T is the temperature, $k_B$ is the Boltzmann constant, $\nu$ is the attempt frequency ($10^{12}$ s$^{-1}$), and the desorption energy barrier can be approximated as the adsorption energy.

For the calculations of the phonon spectrum, the Phonopy code[60] associated with VASP was used. These calculations were performed using the 3x3x1 supercell and finite displacement approaches with the atomic displacement distance of 0.01 Å. The optical properties were calculated based on the TD-HSE06 approach which effectively treats the excitonic effects in 2D materials.[61]

## ACKNOWLEDGMENTS

The Authors thank Mrs. Larisa E. Kistanova for help in creating the graphical illustrations. A.A.K., M.H., and W.C. acknowledge the financial support provided by the Academy of Finland (Grant No. 311934). S.A.Sh. acknowledges the financial support by the Ministry of Science and Higher Education of the Russian Federation (task No. 0784-2020-0027). O.V.P. acknowledge funding of the U. S. National Science Foundation, grant No. CHE-1900510. The authors wish to acknowledge CSC – IT Center for Science, Finland and Peter the Great Saint-Petersburg Polytechnic University Supercomputing Center for computational resources.